\documentclass{article}
\usepackage[round]{natbib}
\usepackage[english]{babel}
\usepackage{amsmath}
\usepackage{amssymb}
\usepackage{amsthm}
\usepackage{mathtools}
\usepackage{physics}
\usepackage{epsfig}
\usepackage{graphics}
\usepackage{bm}
\usepackage{tikz}
\usetikzlibrary{arrows,positioning,shapes.geometric,decorations.markings}
\tikzstyle{squarenode}=[rectangle,draw]
\tikzstyle{littlenode}=[circle,draw,minimum size=0.5cm,font=\small]
\tikzstyle{bigellipse}=[ellipse,draw,x radius=10cm, y radius=5cm]
\tikzset{negate/.style={
            decoration={markings,
            mark= at position 0.5 with {
                  \node[transform shape] (tempnode) {$\Big\Vert$};
                  }
              },
              postaction={decorate}
}
}
\usepackage{xcolor}
\usepackage{tabularx}
\usepackage{multirow}
\usepackage{url}
\usepackage{siunitx,booktabs}
\usepackage{rotating}
\usepackage{caption}
\usepackage{floatrow}
\usepackage{graphicx} 
\usepackage{graphics} 
\usepackage[skip=0.5ex, belowskip=1ex,labelformat=brace,singlelinecheck=off]{subcaption}
\usepackage{pgfplots}
\usepackage{authblk}
\usepackage{etoolbox}

\newcommand{\ind}{\mbox{$\perp\!\!\!\perp$}}

\pgfplotsset{compat=1.11}
\RequirePackage{hyperref}

\usepackage{marvosym}
\usepackage[marvosym]{tikzsymbols}

\usetikzlibrary{arrows,backgrounds,automata,positioning}
\usetikzlibrary{shapes.geometric,decorations.markings}
\usetikzlibrary{decorations.text}
\usetikzlibrary{decorations.pathmorphing}
\tikzstyle{mynode}=[circle,draw,minimum size=0.8cm]
\tikzstyle{myemptynode}=[circle,minimum size=0.8cm]
\tikzstyle{squarenode}=[rectangle,draw]
\tikzstyle{little}=[circle,draw,fill=black,minimum size=0.15cm,inner sep=0pt]
\tikzstyle{littlehollow}=[circle,draw,minimum size=0.15cm,inner sep=0pt]
\tikzset{negate/.style={
            decoration={markings,
            mark= at position 0.5 with {
                  \node[transform shape] (tempnode) {$||$};
                  }
              },
              postaction={decorate}
}
}

\hypersetup{
	colorlinks=true,
	linkcolor=black,
	citecolor=black,
	filecolor=black,
	urlcolor=black,
}
\def\ind{\perp\!\!\!\perp}

\theoremstyle{plain}

\makeatletter
\renewcommand*\env@matrix[1][\arraystretch]{%
  \edef\arraystretch{#1}%
  \hskip -\arraycolsep
  \let\@ifnextchar\new@ifnextchar
  \array{*\c@MaxMatrixCols c}}
\makeatother

\usepackage{soul}

\newcolumntype{Y}{>{\centering\arraybackslash}X}
\newcolumntype{M}{>{\centering\arraybackslash}m}

\title{Interval identification of natural effects in the presence of outcome-related unmeasured confounding}
\author[1]{Marco Doretti\thanks{email: marco.doretti@unifi.it}}
\author[2]{Elena Stanghellini}
\affil[1]{Department of Statistics, Computer Science, and Applications, University of Florence}
\affil[2]{Department of Economics, University of Perugia}

\date{}

\begin{document}
\maketitle

\abstract{With reference to a binary outcome and a binary mediator, we derive identification bounds for natural effects under a reduced set of assumptions. Specifically, no assumptions about confounding are made that involve the outcome; we only assume no unobserved exposure-mediator confounding as well as a condition termed partially constant cross-world dependence (PC-CWD), which poses fewer constraints on the counterfactual probabilities than the usual cross-world independence assumption. The proposed strategy can be used also to achieve interval identification of the total effect, which is no longer point identified under the considered set of assumptions. Our derivations are based on postulating a logistic regression model for the mediator as well as for the outcome. However, in both cases the functional form governing the dependence on the explanatory variables is allowed to be arbitrary, thereby resulting in a semi-parametric approach. To account for sampling variability, we provide delta-method approximations of standard errors in order to build uncertainty intervals from identification bounds. The proposed method is applied to a dataset gathered from a Spanish prospective cohort study. The aim is to evaluate whether the effect of smoking on lung cancer risk is mediated by the onset of pulmonary emphysema.}

%%%%%%%%%%%%%%%%%%%%%%%%%%%%%%%%%

\section{Introduction}\label{sec:intro}
Many research questions involve not only the quantification of the total effect of a treatment/exposure on an outcome of interest, but also the decomposition of this effect into direct and indirect ones, the latter due to possible mediators along the pathway from the treatment to the outcome. In a counterfactual framework, a well-known decomposition involves the so-called \emph{natural} effects. With reference to one mediator, these effects were introduced by~\cite{Pearl2001}, who also provided expressions for their non-parametric identification.

A number of crucial assumptions are required to identify natural effects. Following~\cite{Pearl2001}, most authors specify a sufficient set that  includes the absence of unmeasured exposure-outcome, exposure-mediator and outcome-mediator confounding as well as Cross-World Independence (CWI), a condition on the probabilistic relationship between the mediator and the outcome across two distinct interventional settings. Other authors, in contrast, follow the Sequential Ignorability (SI) approach of~\cite{Imai2010identification}, where identification is reached via an ignorable assignment mechanism that is postulated to hold for both the exposure and the mediator. Both approaches are encoded in the language of Conditional Independence~\citep[CI,][]{Dawid1979}  and are usually deemed equivalent up to probabilistic subtleties~\citep{Forastiereetal2018}.

None of the above assumptions is testable from observed data. The absence of unmeasured confounding could in principle be enforced by randomization or by conditioning upon a suitable set of covariates. However, as it is well-known, randomization is not always feasible, and in observational studies available covariates often fail to completely remove the three kinds of confounding. On the other hand, CWI cannot be fulfilled even in randomized experiments, as it invokes more than one interventional setting at the same time~(\citealt{TchetgenVdW2014}; see also~\citealt{Dawid2000}). In Non-Parametric Structural Equation Models (NPSEMs) with independent error terms~\citep{PearlCausalityBook2009}, it is equivalent to the absence of outcome-mediator confounders affected by the exposure, regardless of whether these confounders are measured or not. Outside the NPSEM framework, however, this is not necessarily the case; in fact, it has been shown that CWI might fail even when outcome-mediator confounding is not present~\citep{AndrewsDidelez2021}. Within the SI paradigm, similar issues arise.

Strategies to partially relax the above assumptions have been put forward within different setups. With specific regard to CWI, these include identifying natural effects under weaker conditions~\citep{robins1992identifiability,Petersenetal2006,TchetgenVdW2014} or particular parametric structures~\citep{DeStavolaetal2015}, but also the definition of alternative estimands like interventional effects~\citep{VdWetal2014,VansteelandtDaniel2017}, which stem from approaches developed outside the counterfactual framework~\citep{Geneletti2007}. \cite{DingVdW2016} provide a lower (upper) bound for the natural direct (indirect) effect without invoking CWI and simultaneously allowing for unmeasured mediator-outcome confounders, though assuming that these are conditionally independent of the exposure given the observed covariates. Other authors explore interval rather than point identification of the natural direct effect in a non-parametric setting where all variables are binary~\citep{RobinsRichardson2010,TchetgenPhiri2014,Kaufmanetal2009}, with subsequent extensions dealing with categorical mediators or outcomes~\citep{Milesetal2017}. Sensitivity analyses have also been carried out within an SI paradigm. These mainly focus on relaxing SI of the mediator, both in a nonparametric~\citep{Sjolander2009} and parametric setting~\citep{Imaietal2010general,Imai2010identification}. In the latter, sensitivity arises from varying the correlation between the error terms of the outcome and mediator regression equation. This logic has been recently extended to the exposure-mediator and exposure-outcome error pairs~\citep{Lindmarketal2018,Lindmark2022}.

In this paper, we provide identification intervals for the natural effects under a restricted set of assumptions when both the mediator and the outcome are binary. We adopt a semi-parametric strategy, in the sense that two logistic regression models are postulated for these variables, but the functional form governing dependence on the explanatory variables is left completely unspecified. We assume no unmeasured exposure-mediator confounding together with an \emph{ad hoc} condition that we term Partially Constant Cross-World Dependence (PC-CWD). We show that this condition poses fewer constraints on the counterfactual probabilities than CWI. Our identification result can be readily used to obtain bounds also for the total effect that, without assumptions on unobserved confounding concerning the outcome, is no longer point identified.

The remainder of the paper is organized as follows. In Section~\ref{sec:pointid}, we review the standard approach leading to point identification of the total effect as well as of natural effects, with specific focus on the case of a binary mediator and a binary outcome. In Section~\ref{sec:interval}, we introduce the PC-CWD condition, including a formal comparison with CWI. Then, our semiparametric bounds are presented. In Section~\ref{sec:est}, we turn our attention to estimation and report formulas to account for sampling variability via the construction of uncertainty intervals, obtained by suitably widening the estimated identification bounds. In Section~\ref{sec:appl}, we apply the proposed approach to data from a prospective cohort study conducted in Spain. Our aim is to evaluate whether the effect of smoking history on lung cancer risk is mediated by the onset of pulmonary emphysema. In Section~\ref{sec:concl}, some concluding remarks are offered.

%%%%%%%%%%%%%%%%%%%%%%%%%%%%%%%%%

\section{Background}\label{sec:pointid}
\subsection{Notation and assumptions}\label{subsec:notass}
We denote the binary outcome by $Y$, the binary mediator by $M$ and the treatment/exposure, that can be of any nature, by $X$. We also introduce the set of observed covariates, $C$. We rely on the counterfactual notation, denoting by $Y(x)$ and $M(x)$ the (random) outcome and mediator under an intervention setting, possibly contrary to the facts, $X$ to the level $x$. Also, we let $Y(x,m)$ be a random variable representing the outcome when $X$ and $M$ are set to $x$ and $m$, respectively.

Throughout the paper we make the usual \emph{consistency} and \emph{composition} assumptions to link observed and counterfactual quantities. Consistency states that, when $X=x$, the counterfactuals $M(x)$ and $Y(x)$ equal the observable variables $M$ and $Y$ and that, conditionally on $X=x$ and $M=m$, $Y(x,m)$ equals $Y$. Composition deals with nested counterfactuals, stating that $Y(x)=Y(x,M(x))$, i.e., that $Y(x)$ can also be seen as the outcome obtained under an hypothetical intervention setting $X$ to $x$ and leaving $M$ to the (random) value it would naturally attain under $X=x$. 

In this setting, the Total Effect (TE) on the outcome (conditionally on $C=c$) of a shift from a reference treatment level, $x^{\star}$, to an active level, $x$, can be measured on various scales. We here consider the log odds-ratio scale~\citep{VdWVansteelandt2010}, that is
\begin{equation}\label{eq:tedef}
\log\textup{OR}_{x,x^{\star}\mid c}^{\textup{TE}} = \log\frac{P(Y(x)=1\mid C=c)/P(Y(x)=0\mid C=c)}{P(Y(x^{\star})=1\mid C=c)/P(Y(x^{\star})=0\mid C=c)}.
\end{equation}
Under the composition assumption, the above effect can be additively decomposed as
\begin{equation}\label{eq:add}
\log\textup{OR}_{x,x^{\star}\mid c}^{\textup{TE}} = \log\textup{OR}_{x,x^{\star}\mid c}^{\textup{NDE}} + \log\textup{OR}_{x,x^{\star}\mid c}^{\textup{NIE}},
\end{equation}
where 
\begin{equation}\label{eq:ndedef}
\log\textup{OR}_{x,x^{\star}\mid c}^{\textup{NDE}} = \log\frac{P(Y(x,M(x^{\star}))=1\mid C=c)/P(Y(x,M(x^{\star}))=0\mid C=c)}{P(Y(x^{\star},M(x^{\star}))=1\mid C=c)/P(Y(x^{\star},M(x^{\star}))=0\mid C=c)}
\end{equation}
is the Natural Direct Effect (NDE) and
\begin{equation}\label{eq:niedef}
\log\textup{OR}_{x,x^{\star}\mid c}^{\textup{NIE}} = \log\frac{P(Y(x,M(x))=1\mid C=c)/P(Y(x,M(x))=0\mid C=c)}{P(Y(x,M(x^{\star}))=1\mid C=c)/P(Y(x,M(x^{\star}))=0\mid C=c)}
\end{equation}
is the Natural Indirect Effect (NIE). Natural effects involve the crossed counterfactual $Y(x,M(x^{\star}))$, which represents the outcome when $X$ is set to $x$ but $M$ is left to the value it would naturally attain under $X=x^{\star}$. 

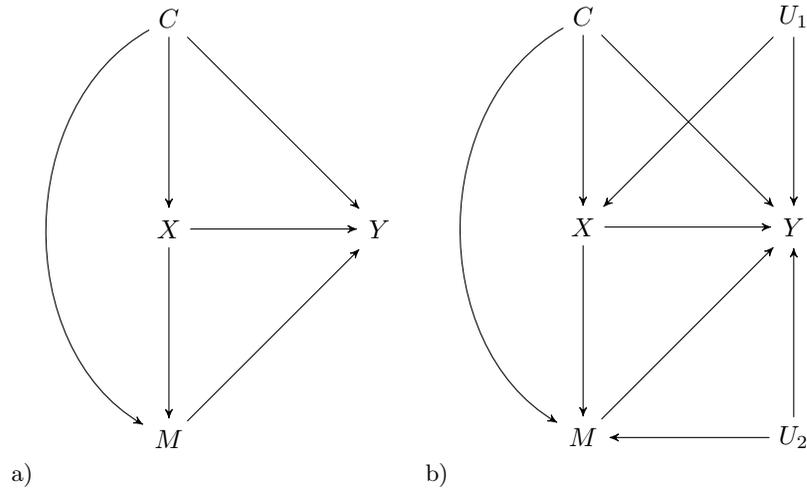
\begin{figure}[tb]
\centering{
\subfloat[][]{\label{fig:mednouc}}
\begin{tikzpicture}[scale=0.3,auto,->,>=stealth',shorten >=1pt,node distance=2.8cm] 
\node (X) {$X$};
\node (Y) [right of=X]{$Y$};
\node (M) [below of=X] {$M$};
\node (C) [above of=X] {$C$};
\draw[->] (X) --node {} (Y); 
\draw[->] (X) --node {} (M); 
%\draw[->] (M) to [out=310, in=330] (Y);
\draw[->] (M) to (Y);
\draw[->] (C) --node {} (X); 
\draw[->] (C) --node {} (Y); 
\draw[->] (C) to [bend right=60] (M);
\end{tikzpicture}\quad
\subfloat[][]{\label{fig:meduc}}
\begin{tikzpicture}[scale=0.3,auto,->,>=stealth',shorten >=1pt,node distance=2.8cm] 
\node (X) {$X$};
\node (Y) [right of=X]{$Y$};
\node (M) [below of=X] {$M$};
\node (C) [above of=X] {$C$};
\node (U1) [above of=Y] {$U_1$};
\node (U2) [below of=Y] {$U_2$};
\draw[->] (X) --node {} (Y); 
\draw[->] (X) --node {} (M); 
%\draw[->] (M) to [out=310, in=330] (Y);
\draw[->] (M) to (Y);
\draw[->] (C) --node {} (X); 
\draw[->] (C) --node {} (Y); 
\draw[->] (C) to [bend right=60] (M);
\draw[->] (U1) --node {} (X); 
\draw[->] (U1) --node {} (Y);
\draw[->] (U2) --node {} (M); 
\draw[->] (U2) --node {} (Y);
\end{tikzpicture}
}
\caption{Mediation scheme for $(X,M,Y)$ with measured covariates $C$ and: a) no unmeasured exposure-mediator, exposure-outcome and mediator-outcome confounders, b) unmeasured exposure-outcome ($U_1$) and mediator-outcome ($U_2$) confounders. In b), bidirected arrows between $C$, $U_1$ and $U_2$ are omitted but possible.}
\label{fig:med}
\end{figure}

The total effect is identified assuming \emph{(i*)} $Y(x)\ind X\mid C=c$, which corresponds to no unmeasured exposure-outcome confounding. In contrast, sufficient conditions for identification of natural effects are: \emph{(i)} $Y(x,m) \ind X \mid C=c$, \emph{(ii)} $Y(x,m) \ind M \mid (X=x,C=c)$, \emph{(iii)} $M(x) \ind X \mid C=c$, and \emph{(iv)} $Y(x,m) \ind M(x^{\star}) \mid C=c$. While \emph{(iv)} encodes CWI, \emph{(i)}-\emph{(iii)} are related to confounding. In particular, \emph{(i)} is a modified version of \emph{(i*)} tailored to the mediation context, whereas \emph{(ii)} and \emph{(iii)} respectively correspond to no unmeasured outcome-mediator and exposure-mediator confounding; see the corresponding causal diagram~\citep[][Chapter 3]{PearlCausalityBook2009} in Figure~\ref{fig:mednouc}.

All the above CI statements are intended to hold for every level $x$, $x^{\star}$ and $m$ (with $x\neq x^{\star}$), which makes them not testable from observed data. Assumptions \emph{(i)-(iii)} can in principle be weakened by diversifying the covariate set with the kind of confounding (exposure-outcome, mediator-outcome and exposure-mediator) and/or considering identification strategies other than covariate adjustment; see~\cite{Pearl2014} for details. However, we here leave them as stated for the sake of simplicity, and we defer further comments to the discussion. 

\subsection{Point identification}\label{subsec:pident}
Under assumption \emph{(i*)}, for any level $x$ and $y$ it holds that
\begin{equation}\label{eq:identte}
P(Y(x)=y\mid C=c)=P(Y=y\mid X=x,C=c).
\end{equation}
Thus, point identification of the TE in~\eqref{eq:tedef} is achieved by plugging-in the observed probabilities, be it in a non-parametric or parametric setting. With regard to natural effects, assumptions \emph{(i)-(iv)} allow non-parametric identification of the crossed counterfactual probability $P(Y(x,M(x^{\star}))=y\mid C=c)$ via Pearl's mediation formula
\[
P(Y(x,M(x^{\star}))=y\mid C=c) = \sum_m P( Y=y \mid X=x,M=m,C=c)P(M=m\mid X=x^{\star},C=c)
\]
\citep{Pearl2001,Pearl2010med}. Similarly, when $x^{\star}=x$ we have
\[
\begin{split}
P(Y(x)=y\mid C=c) &= \sum_m P( Y=y \mid X=x,M=m,C=c)P(M=m\mid X=x,C=c) \\
&=P( Y=y \mid X=x,C=c),
\end{split}
\]
coherently with~\eqref{eq:identte}. Notice that this requires the CI statement \emph{(v)} $Y(x,m)\ind M(x)\mid C=c$, sometimes termed Single-World Independence (SWI), to hold. Such a statement is implied by assumptions \emph{(i)-(iii)}, see Theorem 1 of~\cite{RobinsRichardson2010}. As a consequence, \emph{(i)-(iv)} allow to point identify the natural effects~\eqref{eq:ndedef} and~\eqref{eq:niedef} with a plug-in strategy in the same way as the total effect.

Within a semi-parametric framework, one could specify the outcome and mediator logistic models
\begin{equation}\label{py}
\textup{logit}\{P(Y=1 \mid X=x,M=m,C=c)\}  = r_{\beta}(x,m;c)
\end{equation}
and
\begin{equation}\label{pm}
\textup{logit}\{P(M=1 \mid X=x,C=c)\}=  r_{\gamma}(x;c),
\end{equation}
where the functional form of $r_{\beta}(x,m;c)$ and $r_{\gamma}(x;c)$ is arbitrary. In this setting, it is easy to extend the fully parametric strategy considered in~\cite{Dorettietal2022} and identify the natural effect logistic model with
\begin{equation}\label{eq:nem}
\log\frac{P(Y(x,M(x^*))=1 \mid C=c)}{P(Y(x,M(x^*))=0\mid C=c)} = r_{\beta}(x,0;c)+\log\frac{1+\exp\{g_1(x,x^{\star};c)\}}{1+\exp\{g_0(x,x^{\star};c)\}},
\end{equation}
where
\begin{equation} \label{gcontra}
\begin{split}
g_{y}(x, x^{\star};c) &= \textup{logit}\{P(M(x^{\star})=1 \mid Y(x,M(x^{\star}))=y,C=c)\} \\
&= \log\frac{P(Y=y\mid M=1,X=x,C=c)}{P(Y=y\mid M=0,X=x,C=c)} + \log\frac{P(M=1\mid X=x^{\star},C=c)}{P(M=0\mid X=x^{\star},C=c)} \\
&= y\{r_{\beta}(x,1;c)-r_{\beta}(x,0;c)\}+\log\frac{1+\exp\{r_{\beta}(x,0;c)\}}{1+\exp\{r_{\beta}(x,1;c)\}}+r_{\gamma}(x^{\star};c)
\end{split}
\end{equation}
(see Appendix~\ref{app:proofnem} for details). Under SWI, the same developments lead to
\begin{equation}\label{eq:margmod}
\log\frac{P(Y(x)=1 \mid C=c)}{P(Y(x)=0\mid C=c)} = r_{\beta}(x,0;c)+\log\frac{1+\exp\{g_1(x,x;c)\}}{1+\exp\{g_0(x,x;c)\}},
\end{equation}
where
\[
\begin{split}
g_y(x,x;c) &= \textup{logit}\{P(M=1 \mid Y=y,X=x,C=c)\} \\
&= \log\frac{P(Y=y\mid M=1,X=x,C=c)}{P(Y=y\mid M=0,X=x,C=c)} + \log\frac{P(M=1\mid X=x,C=c)}{P(M=0\mid X=x,C=c)} \\
&= y\{r_{\beta}(x,1;c)-r_{\beta}(x,0;c)\}+\log\frac{1+\exp\{r_{\beta}(x,0;c)\}}{1+\exp\{r_{\beta}(x,1;c)\}}+r_{\gamma}(x;c)
\end{split}
\]
is a special case of~\eqref{gcontra} in which the treatment level is the same in both the outcome and mediator model part.

From~\eqref{eq:nem} and~\eqref{eq:margmod}, it is immediate to obtain the expressions for point identification of log odds-ratio natural effects, that is,
\[
\log\textup{OR}^{\textup{NDE}}_{x,x^{\star}\mid c}=  r_{\beta}(x,0;c)-r_{\beta}(x^{\star},0;c) + \log\frac{1+\exp\{g_1(x,x^{\star};c)\}}{1+\exp\{g_0(x,x^{\star};c)\}} - \log\frac{1+\exp\{g_1(x^{\star},x^{\star};c)\}}{1+\exp\{g_0(x^{\star},x^{\star};c)\}}
\]
and 
\[
\log\textup{OR}^{\textup{NIE}}_{x,x^{\star}\mid c}= \log\frac{1+\exp\{g_1(x,x;c)\}}{1+\exp\{g_0(x,x;c)\}} - \log\frac{1+\exp\{g_1(x,x^{\star};c)\}}{1+\exp\{g_0(x,x^{\star};c)\}}.
\]
The sum of the two above effects returns the TE
\begin{equation}\label{eq:logorte}
\log\textup{OR}_{x,x^{\star}\mid c}^{\textup{TE}} = r_{\beta}(x,0;c) - r_{\beta}(x^{\star},0;c) + \log\frac{1+\exp\{g_1(x,x;c)\}}{1+\exp\{g_0(x,x;c)\}} - \log\frac{1+\exp\{g_1(x^{\star},x^{\star};c)\}}{1+\exp\{g_0(x^{\star},x^{\star};c)\}},
\end{equation}
in line with recent results on marginal and conditional effects in logistic regression outside the counterfactual framework; see~\cite{StanghelliniDoretti2019} and~\cite{Raggietal2021}.

%%%%%%%%%%%%%%%%%%%%%%%%%%%%%%%%%%%%%%

\section{Interval identification}\label{sec:interval}
We now remove the outcome-related confounding assumptions \emph{(i)} and \emph{(ii)} allowing for unmeasured confounders of the exposure-outcome and mediator-outcome relations; see the diagram in Figure~\ref{fig:meduc}. Moreover, we replace CWI \emph{(iv)} with the PC-CWD condition introduced in Section~\ref{sec:intro}, which reads as
\begin{equation}\label{eq:pccwd}\tag{\emph{iv*}}
P(Y(x,m)=y\mid M(x^{\star})=m,C=c) = P(Y(x,m)=y\mid M(x)=m,C=c).
\end{equation}
In words,~\emph{\eqref{eq:pccwd}} states that, within covariate levels, the probability distribution of $Y(x,m)$ conditional on the event $M(x^{\star})=m$ equals the one conditional on the event $M(x)=m$. However, no specific values for these conditional distributions are posited, nor any claim is made about the impact of conditioning on the events $M(x^{\star})=1-m$ and $M(x)=1-m$. Thus, the pairwise stochastic dependence relationships between $Y(x,m)$ and both $M(x)$ and $M(x^{\star})$ remain arbitrary, with no CI statements implied for them. For this reason, we refer to~\eqref{eq:pccwd} as to ``Partially Constant Cross-World Dependence'', where the expression ``Partially Constant'' reflects the fact that only one of the two conditional cross-world distributions of $Y(x,m)$ is subject to some kind of constraint (that is, of being equal to the corresponding single-world distribution).

With respect to the standard identification approach of Section~\ref{sec:pointid}, where CWI is assumed and SWI is implied by \emph{(i)-(iii)}, PC-CWD poses a considerably lesser amount of restrictions on the conditional probabilities of counterfactual outcomes. We formally show this in Table~\ref{tab:comp}, where we list, for both settings, the probabilities that have to be equal to $P(Y(x,m)=1\mid M(x)=m, C=c)$ in consequence of the assumptions made. As mentioned before, the latter probability is unconstrained; a coherent parametrization is given by
\begin{equation}\label{eq:condlogit}
%\begin{split}
\textup{logit}\{P(Y(x,m) = 1\mid M(x)=m,C=c)\} = \psi + r_{\beta}(x,m;c),
%\end{split}
\end{equation}
where $r_{\beta}(x,m;c)$ is the predictor of the postulated semi-parametric outcome model~\eqref{py} and $\psi$ is an arbitrary real-valued additive shift. Such a parametrization forms, together with~\eqref{eq:pccwd}, the basis of our interval identification strategy. It is important to remark that, since \emph{(i)} and \emph{(ii)} are no longer invoked, there is no guarantee that 
\[
\textup{logit}\{P(Y(x,m) = 1\mid C=c)\} = r_{\beta}(x,m;c)
\]
as in the standard identification approach enhanced with model~\eqref{py}. Therefore, $\psi=0$ does not necessarily correspond to the SWI assumption holding.

\begin{table}[tb]
\begin{tabularx}{0.7\textwidth}{lYY}
\toprule
 & \textbf{CWI + SWI} & \textbf{PC-CWD} \\
 \midrule
$P(Y(x,m)=1\mid M(x)=1-m, C=c)$ &  \checkmark &  \\
$P(Y(x,m)=1\mid C=c)$ &  \checkmark &  \\
$P(Y(x,m)=1\mid M(x^{\star})=m, C=c)$ & \checkmark & \checkmark \\
$P(Y(x,m)=1\mid M(x^{\star})=1-m, C=c)$ &  \checkmark &  \\
\bottomrule
\end{tabularx}
\caption{Counterfactual probabilities that are constrained to be equal to $P(Y(x,m)=1\mid M(x)=m, C=c)$ under CWI (paired with SWI) and PC-CWD.}\label{tab:comp}
\end{table}

The identification strategy outlined in Section~\ref{subsec:pident} (and Appendix~\ref{app:proofnem}) can be implemented also with the reduced assumption set considered in this section. Specifically, under assumptions \emph{(iii)} and~\eqref{eq:pccwd} it is possible to exploit the parametrization in~\eqref{eq:condlogit} in order to update the semi-parametric expression of the natural effect model with
\begin{equation}\label{eq:nemsens}
\text{logit}\{P(Y(x,M(x^{\star}))=1\mid C=c)\} = O(\psi,x,0;c) + \log \frac{1+\exp\{\tilde{g}_1(\psi,x,x^{\star};c)\}}{1+\exp\{\tilde{g}_0(\psi,x,x^{\star};c)\}},
\end{equation}
where $O(\psi,x,m;c)$ is a compact form for the right-hand side in~\eqref{eq:condlogit} and
\begin{equation} \label{eq:gtilde}
\begin{split}
\tilde{g}_y(\psi,x, x^{\star};c) &= y\{O(\psi,x,1;c)-O(\psi,x,0;c)\}+\log\frac{1+\exp\{O(\psi,x,0;c)\}}{1+\exp\{O(\psi,x,1;c)\}}+r_{\gamma}(x^{\star};c) \\
&= y\{r_{\beta}(x,1;c)-r_{\beta}(x,0;c)\}+\log\frac{1+\exp\{O(\psi,x,0;c)\}}{1+\exp\{O(\psi,x,1;c)\}}+r_{\gamma}(x^{\star};c)
\end{split}
\end{equation}
($y=0,1$) is a modified version of $g_y(x,x^{\star};c)$ in~\eqref{gcontra}. In parallel with Section~\ref{subsec:pident}, the corresponding expression for the single-world probability $P(Y(x)=y\mid C=c)$ is
\[
\text{logit}\{P(Y(x,M(x))=1\mid C=c)\} = O(\psi,x,0;c) + \log \frac{1+\exp\{\tilde{g}_1(\psi,x,x;c)\}}{1+\exp\{\tilde{g}_0(\psi,x,x;c)\}}.
\]
Consequently, the updated semi-parametric formulas for natural effects are
\begin{equation}\label{eq:ndepsi}
\log\textup{OR}^{\textup{NDE}}_{x,x^{\star}\mid c} = r_{\beta}(x,0;c)-r_{\beta}(x^{\star},0;c)+\log \frac{1+\exp\{\tilde{g}_1(\psi,x,x^{\star};c)\}}{1+\exp\{\tilde{g}_0(\psi,x,x^{\star};c)\}} - \log \frac{1+\exp\{\tilde{g}_1(\psi,x^{\star},x^{\star};c)\}}{1+\exp\{\tilde{g}_0(\psi,x^{\star},x^{\star};c)\}}
\end{equation}
and
\begin{equation}\label{eq:niepsi}
\log\textup{OR}^{\textup{NIE}}_{x,x^{\star}\mid c} = \log \frac{1+\exp\{\tilde{g}_1(\psi,x,x;c)\}}{1+\exp\{\tilde{g}_0(\psi,x,x;c)\}} - \log \frac{1+\exp\{\tilde{g}_1(\psi,x,x^{\star};c)\}}{1+\exp\{\tilde{g}_0(\psi,x,x^{\star};c)\}}.
\end{equation}

Importantly, in~\eqref{eq:gtilde} the impact of the arbitrary shift $\psi$ is confined to the logarithmic term. Thus, letting
\[
\delta_{\beta}(x;c)\equiv r_{\beta}(x,1;c)-r_{\beta}(x,0;c)
\]
be the log odds-ratio scale effect of the mediator on the outcome, evaluated at $x$, we have
\[
\tilde{g}_1(\psi,x,x^{\star};c) = \tilde{g}_0(\psi,x,x^{\star};c) + \delta_{\beta}(x;c),
\]
so that the argument of the logarithmic term in~\eqref{eq:nemsens} can be expressed, after simple algebra, by means of the straight line equation
\begin{equation}\label{eq:fpsi}
\frac{1+\exp\{\tilde{g}_1(\psi,x,x^{\star};c)\}}{1+\exp\{\tilde{g}_0(\psi,x,x^{\star};c)\}} =\exp\{\delta_{\beta}(x;c)\} + [1-\exp\{\delta_{\beta}(x;c)\}]p_{x,x^{\star}\mid c}^{(\psi)}.
\end{equation}
In the above, $\exp\{\delta_{\beta}(x;c)\}$ is the intercept, the slope $1-\exp\{\delta_{\beta}(x;c)\}$ takes the opposite sign of $\delta_{\beta}(x;c)$ and
\begin{equation}\label{eq:psens}
\begin{split}
p_{x,x^{\star}\mid c}^{(\psi)} &= P(M(x^{\star})=0\mid Y(x,M(x^{\star}))=0,C=c) \\
&= 1/[1+\exp\{\tilde{g}_0(\psi,x,x^{\star};c)\}]
\end{split}
\end{equation}
 can be regarded as a sensitivity parameter lying between 0 and 1. In fact, it is possible to show that $p^{(\psi)}_{x,x^{\star}\mid c}$ varies in a narrower interval, that is,
\begin{equation}\label{eq:boundsp1}
\frac{1}{1+\exp\{r_{\gamma}(x^{\star};c)\}} \;\leq\; p^{(\psi)}_{x,x^{\star}\mid c} \;\leq\; \frac{\exp\{\delta_{\beta}(x;c)\}}{\exp\{\delta_{\beta}(x;c)\}+\exp\{r_{\gamma}(x^{\star};c)\}} 
\end{equation}
when $\delta_{\beta}(x;c)>0 $ and
\begin{equation}\label{eq:boundsp2}
\frac{\exp\{\delta_{\beta}(x;c)\}}{\exp\{\delta_{\beta}(x;c)\}+\exp\{r_{\gamma}(x^{\star};c)\}}  \;\leq\; p^{(\psi)}_{x,x^{\star}\mid c} \;\leq\; \frac{1}{1+\exp\{r_{\gamma}(x^{\star};c)\}} 
\end{equation}
when $\delta_{\beta}(x;c)<0$; see Appendix~\ref{app:boundsp} for a formal proof. Performing a sensitivity analysis on the naturally bounded quantity $p^{(\psi)}_{x,x^{\star}\mid c}$  - rather than on $\psi$ - allows to bound the left-hand side of~\eqref{eq:fpsi} as 
\[
\ell_{x,x^{\star}\mid c} \;\leq\; \frac{1+\exp\{\tilde{g}_1(\psi,x,x^{\star};c)\}}{1+\exp\{\tilde{g}_0(\psi,x,x^{\star};c)\}} \;\leq\;  u_{x,x^{\star}\mid c},
\]
where
\[
\ell_{x,x^{\star}\mid c}\equiv\frac{\exp\{\delta_{\beta}(x;c)\}[1+\exp\{r_{\gamma}(x^{\star};c)\}]}{\exp\{\delta_{\beta}(x;c)\}+\exp\{r_{\gamma}(x^{\star};c)\}} \qquad \qquad
u_{x,x^{\star}\mid c} \equiv \frac{1+\exp\{\delta_{\beta}(x;c)\}\exp\{r_{\gamma}(x^{\star};c)\}}{1+\exp\{r_{\gamma}(x^{\star};c)\}}.
\]
Such inequalities immediately follow from~\eqref{eq:fpsi}-\eqref{eq:boundsp2}, and they hold whatever the sign of $\delta_{\beta}(x;c)$ is. Since this result holds for every level pair, including $(x,x)$ and $(x^{\star},x^{\star})$, it is possible to embed it in~\eqref{eq:ndepsi} and~\eqref{eq:niepsi} in order to obtain identification bounds for log odds-ratio natural effects. In practice, we have
\begin{equation}\label{eq:ideint}
\openup 5\jot
\begin{split}
\log \ell^{\textup{NDE}}_{x,x^{\star}\mid c} \;\leq\;& \log\textup{OR}^{\textup{NDE}}_{x,x^{\star}\mid c} \;\leq\;  \log u^{\textup{NDE}}_{x,x^{\star}\mid c} \\
\log \ell^{\textup{NIE}}_{x,x^{\star}\mid c} \;\leq\;& \log\textup{OR}^{\textup{NIE}}_{x,x^{\star}\mid c} \;\leq\;  \log u^{\textup{NIE}}_{x,x^{\star}\mid c},
\end{split}
\end{equation}
where the odds-ratio scale bounds for the direct and indirect effect are respectively given by
\[
\openup 5\jot
\begin{split}
\ell^{\textup{NDE}}_{x,x^{\star}\mid c} &=  \frac{\exp\{r_{\beta}(x,0;c)\}}{\exp\{r_{\beta}(x^{\star},0;c)\}}\cdot \frac{\ell_{x,x^{\star}\mid c}}{u_{x^{\star},x^{\star}\mid c}} \\
&= \frac{\exp\{r_{\beta}(x,0;c)\}}{\exp\{r_{\beta}(x^{\star},0;c)\}}\cdot\frac{[1+\exp\{r_{\gamma}(x^{\star};c)\}]^2}{[1+\exp\{r_{\gamma}(x^{\star};c)-\delta_{\beta}(x;c)\}][1+\exp\{r_{\gamma}(x^{\star};c)+\delta_{\beta}(x^{\star};c)\}]} \\
u^{\textup{NDE}}_{x,x^{\star}\mid c} &= \frac{\exp\{r_{\beta}(x,0;c)\}}{\exp\{r_{\beta}(x^{\star},0;c)\}}\cdot \frac{u_{x,x^{\star}\mid c}}{\ell_{x^{\star},x^{\star}\mid c}}\\
&= \frac{\exp\{r_{\beta}(x,0;c)\}}{\exp\{r_{\beta}(x^{\star},0;c)\}}\cdot\frac{[1+\exp\{r_{\gamma}(x^{\star};c)+\delta_{\beta}(x;c)\}][1+\exp\{r_{\gamma}(x^{\star};c)-\delta_{\beta}(x^{\star};c)\}]}{[1+\exp\{r_{\gamma}(x^{\star};c)\}]^2}
\end{split}
\]
and
\[
\openup 5\jot
\begin{split}
\ell^{\textup{NIE}}_{x,x^{\star}\mid c} &=  \frac{\ell_{x,x\mid c}}{u_{x,x^{\star}\mid c}}= \frac{[1+\exp\{r_{\gamma}(x;c)\}][1+\exp\{r_{\gamma}(x^{\star},c)\}]}{[1+\exp\{r_{\gamma}(x;c)-\delta_{\beta}(x;c)\}][1+\exp\{r_{\gamma}(x^{\star};c)+\delta_{\beta}(x;c)\}]} \\
u^{\textup{NIE}}_{x,x^{\star}\mid c} &= \frac{u_{x,x\mid c}}{\ell_{x,x^{\star}\mid c}} = \frac{[1+\exp\{r_{\gamma}(x;c)+\delta_{\beta}(x;c)\}][1+\exp\{r_{\gamma}(x^{\star};c)-\delta_{\beta}(x;c)\}]}{[1+\exp\{r_{\gamma}(x;c)\}][1+\exp\{r_{\gamma}(x^{\star},c)\}]}.
\end{split}
\]
Identification bounds for the log odds-ratio TE, which is not point-identified in the absence of assumptions \emph{(i)} and \emph{(ii)}, can be readily obtained by exploiting the additive relationship in~\eqref{eq:add}. The resulting expression is
\begin{equation}\label{eq:ideintte}
\log \ell^{\textup{TE}}_{x,x^{\star}\mid c} \;\leq\; \log\textup{OR}^{\textup{TE}}_{x,x^{\star}\mid c} \;\leq\;  \log u^{\textup{TE}}_{x,x^{\star}\mid c},\\
\end{equation}
 where 
 \[
 \openup 5\jot
 \begin{split}
\log \ell^{\textup{TE}}_{x,x^{\star}\mid c} &= \log \ell^{\textup{NDE}}_{x,x^{\star}\mid c} + \log \ell^{\textup{NIE}}_{x,x^{\star}\mid c} \\
\log u^{\textup{TE}}_{x,x^{\star}\mid c} &= \log u^{\textup{NDE}}_{x,x^{\star}\mid c} + \log u^{\textup{NIE}}_{x,x^{\star}\mid c}.
 \end{split}
 \]

%%%%%%%%%%%%%%%%%%%%%%%%%%%%%%%%%%%%%%

\section{Uncertainty intervals}\label{sec:est}
The identification bounds in~\eqref{eq:ideint} and~\eqref{eq:ideintte} are univocally determined by the six-dimensional vector of predictors
\[
\bm{\theta}_{x,x^{\star}\mid c}=\bigl(r_{\beta}(x,0;c),r_{\beta}(x^{\star},0;c),r_{\beta}(x,1;c),r_{\beta}(x^{\star},1;c),r_{\gamma}(x;c),r_{\gamma}(x^{\star};c)\bigr)^\top.
\]
Thus, their finite-sample estimation can be straightforwardly performed by plugging-in the elements of the $\hat{\bm{\theta}}_{x,x^{\star}\mid c}$ estimate everywhere in the formulae. In order to account for sampling variability, it is possible to build uncertainty intervals~\citep{Vansteelandtetal2006}. Although originally introduced to deal with non-ignorable missingness, these have been subsequently adapted for sensitivity analyses in other contexts including selection models~\citep{Genbacketal2015} and causal inference with unobserved confounding, both within~\citep{Lindmarketal2018} and outside~\citep{GenbackdeLuna2019} a mediation context. In our framework, $100\cdot (1-\alpha)$\% point-wise uncertainty intervals for the log odds-ratio effects are given by
\begin{equation}\label{eq:ui}
\openup 5\jot
\begin{split}
\Biggl[\log \widehat{\ell^{\textup{NDE}}_{x,x^{\star}\mid c}} - z_{\alpha/2}\cdot \textup{se}\Bigl(\log \widehat{\ell^{\textup{NDE}}_{x,x^{\star}\mid c}}  \Bigr) \;\; &; \;\;  \log \widehat{u^{\textup{NDE}}_{x,x^{\star}\mid c}} + z_{\alpha/2}\cdot \textup{se}\Bigl(\log \widehat{u^{\textup{NDE}}_{x,x^{\star}\mid c}} \Bigr) \Biggr] \\
\Biggl[\log \widehat{\ell^{\textup{NIE}}_{x,x^{\star}\mid c}} - z_{\alpha/2}\cdot \textup{se}\Bigl(\log \widehat{\ell^{\textup{NIE}}_{x,x^{\star}\mid c}}  \Bigr) \;\; &; \;\;  \log \widehat{u^{\textup{NIE}}_{x,x^{\star}\mid c}} + z_{\alpha/2}\cdot \textup{se}\Bigl(\log \widehat{u^{\textup{NIE}}_{x,x^{\star}\mid c}} \Bigr) \Biggr] \\
\Biggl[\log \widehat{\ell^{\textup{TE}}_{x,x^{\star}\mid c}} - z_{\alpha/2}\cdot \textup{se}\Bigl(\log \widehat{\ell^{\textup{TE}}_{x,x^{\star}\mid c}}  \Bigr) \;\; &; \;\;  \log \widehat{u^{\textup{TE}}_{x,x^{\star}\mid c}} + z_{\alpha/2}\cdot \textup{se}\Bigl(\log \widehat{u^{\textup{TE}}_{x,x^{\star}\mid c}} \Bigr) \Biggr], \\
\end{split}
\end{equation}
where $\textup{se}(\cdot)$ denotes the standard error of the corresponding estimator and $-z_{\alpha/2}$ is the standard normal $\alpha/2$-quantile. 

Approximate expressions for the standard errors in~\eqref{eq:ui} can be obtained via the Delta method~\citep{Oehlert1992}. To this end, we denote by $\bm{\Sigma}_{x,x^{\star}\mid c}$ the variance-covariance matrix of the $\hat{\bm{\theta}}_{x,x^{\star}\mid c}$ estimator, and we assume one is able to reconstruct such a matrix. For instance, when logistic regression models are postulated for the outcome and the mediator, the components of $\hat{\bm{\theta}}_{x,x^{\star}\mid c}$ are linear combinations of the regression coefficient estimators, so it is immediate to obtain $\bm{\Sigma}_{x,x^{\star}\mid c}$ from the variance/covariance matrices of the latter. In this setting, it is convenient to define the vector
\[
\bm{\tau}_{x,x^{\star}\mid c} = \biggl(\log\ell_{x,x^{\star}\mid c}^{\textup{NDE}},\log u_{x,x^{\star}\mid c}^{\textup{NDE}},\log\ell_{x,x^{\star}\mid c}^{\textup{NIE}},\log u_{x,x^{\star}\mid c}^{\textup{NIE}} \biggr)^\top,
\]
with $\hat{\bm{\tau}}_{x,x^{\star}\mid c}$ denoting the corresponding estimator. Then, the variance-covariance matrix $V(\hat{\bm{\tau}}_{x,x^{\star}\mid c})$ can be approximated by
\begin{equation}\label{eq:v0}
V_0(\hat{\bm{\tau}}_{x,x^{\star}\mid c}) = \bm{D}_{x,x^{\star}\mid c}^\top \bm{\Sigma}_{x,x^{\star}\mid c}\bm{D}_{x,x^{\star}\mid c},
\end{equation}
where $\bm{D}_{x,x^{\star}\mid c}=\partial \bm{\tau}^\top_{x,x^{\star}\mid c}/\partial \bm{\theta}_{x,x^{\star}\mid c}$ is a derivative matrix defined in Appendix~\ref{app:delta}. The elements
\[
V_0(\hat{\bm{\tau}}_{x,x^{\star}\mid c})_{[1,3]}\approx\textup{Cov}\Bigl(\log \widehat{\ell^{\textup{NDE}}_{x,x^{\star}\mid c}},\log \widehat{\ell^{\textup{NIE}}_{x,x^{\star}\mid c}}\Bigr)\qquad\quad \text{and} \quad\qquad V_0(\hat{\bm{\tau}}_{x,x^{\star}\mid c})_{[2,4]}\approx \textup{Cov}\Bigl(\log \widehat{u^{\textup{NDE}}_{x,x^{\star}\mid c}},\log \widehat{u^{\textup{NIE}}_{x,x^{\star}\mid c}}\Bigr) 
\]
are needed to compute the variances of the log odds-ratio TE estimator bounds, that is,
\[
V\Bigl(\log \widehat{\ell^{\textup{TE}}_{x,x^{\star}\mid c}}\Bigr) = V\Bigl(\log \widehat{\ell^{\textup{NDE}}_{x,x^{\star}\mid c}}\Bigr) + V\Bigl(\log \widehat{\ell^{\textup{NIE}}_{x,x^{\star}\mid c}}\Bigr) + 2\textup{Cov}\Bigl(\log \widehat{\ell^{\textup{NDE}}_{x,x^{\star}\mid c}},\log \widehat{\ell^{\textup{NIE}}_{x,x^{\star}\mid c}}\Bigr)
\]
and
\[
V\Bigl(\log \widehat{u^{\textup{TE}}_{x,x^{\star}\mid c}}\Bigr) = V\Bigl(\log \widehat{u^{\textup{NDE}}_{x,x^{\star}\mid c}}\Bigr) + V\Bigl(\log \widehat{u^{\textup{NIE}}_{x,x^{\star}\mid c}}\Bigr) + 2\textup{Cov}\Bigl(\log \widehat{u^{\textup{NDE}}_{x,x^{\star}\mid c}},\log \widehat{u^{\textup{NIE}}_{x,x^{\star}\mid c}}\Bigr).
\]
Clearly, an estimate of $V_0(\hat{\bm{\tau}}_{x,x^{\star}\mid c})$, which standard errors can be extracted from, is obtained by replacing $\bm{\Sigma}_{x,x^{\star}\mid c}$ and $\bm{D}_{x,x^{\star}\mid c}$ in its expression with the corresponding finite-sample estimates $\hat{\bm{\Sigma}}_{x,x^{\star}\mid c}$ and $\hat{\bm{D}}_{x,x^{\star}\mid c}$.

%%%%%%%%%%%%%%%%%%%%%%%%%%%

\section{Application}\label{sec:appl}
The interval identification strategy is applied to data collected within the lung cancer screening program implemented by Clinica Universidad de Navarra (CUN, located in Spain) between September 2000 and December 2016~\citep{Gonzalezetal2019}. The resulting prospective cohort study included more than 3000 over-40-aged heavy smokers (cumulative tobacco exposure of at least 10 pack-years) without lung cancer symptoms or other cancer diagnoses within 5 years from baseline. The original aim was investigating the impact of different phenotypes of emphysema, which was found to be a very strong predictor of lung cancer onset~\citep{deTorresetal2007,Wilsonetal2008}, even after adjusting for confounding factors~\citep{Turneretal2007,Lietal2011,Zulueta2012}. Clearly, the latter include tobacco exposure, which is a well-known cause of lung cancer as well as of other pulmonary diseases like emphysema~\citep{Foreyetal2011,TubioPerezetal2020}.

In the above setting, it is worth to investigate whether emphysema, $M$, acts as a mediator on the casual pathway from smoking intensity, $X$, to lung cancer onset $Y$ (1=yes, 0=no). We here consider such a mediation scheme without distinguishing among emphysema types, thereby dealing with a binary mediator $M$ (1=presence of any kind of emphysema, 0=otherwise). In contrast, the exposure $X$ is a continuous variable measured in pack-years. In addition to $X$, $M$ and $Y$, the available covariates in the CUN dataset are Age, Gender (1=male, 0=female) and Body Mass Index (BMI). After preliminary data cleansing, the final dataset includes 3270 individuals; some descriptive statistics are reported in Table~\ref{tab:descr}.

\begin{table}[tb]
\centering
\begin{subtable}[T]{0.45\textwidth}
\begin{tabularx}{\linewidth}{lYYY}
\toprule
 & $X$ & Age & BMI \\
 \cmidrule{2-4}  
Min. & 10 & 40 & 12.551 \\ 
1st Qu. & 20.975 & 49 & 24.509 \\ 
Median & 32 & 55 & 27.213 \\ 
Mean & 36.933 & 56.169 & 27.564 \\ 
3rd Qu. & 46 & 62 & 30.116 \\ 
Max. & 172.300 & 87 & 52.202 \\ 
SD & 21.521 & 9.336 & 4.443 \\
 \midrule
 & $M$ & Gender & $Y$  \\ 
 \cmidrule{2-4}
 0 & 0.755 & 0.275 & 0.980  \\ 
 1 & 0.245 & 0.725 &  0.020 \\ 
\bottomrule
\end{tabularx}
\captionsetup{justification=centering}
\caption{Descriptive statistics}
\label{tab:descr}
\end{subtable}
\hspace*{0.2cm}%
\begin{subtable}[T]{0.45\textwidth}
\begin{tabularx}{\linewidth}{lYYY}
\toprule
{\bf Mediator} & est. & s.e. & $p$-value \\
 \cmidrule{2-4} 
%\multicolumn{4}{l}{\hspace{0.8cm} {\bf Mediator}} \\
 Intercept & 0.418 & 0.296 & 0.157 \\ 
 $X$ & 0.017 & 0.002 & 0.000 \\ 
  BMI & -0.098 & 0.012 & 0.000 \\ 
  Gender & 0.595 & 0.114 & 0.000 \\
\midrule
{\bf Outcome} & est. & s.e. & $p$-value \\
 \cmidrule{2-4} 
  Intercept & -3.925 & 0.899 & 0.000 \\  
  $X$ & 0.020 & 0.004 & 0.000 \\
  $M$ & 1.250 & 0.264 & 0.000 \\ 
  BMI & -0.064 & 0.034 & 0.062 \\ 
  Gender & 0.587 & 0.376 & 0.118 \\ 
\bottomrule
\end{tabularx}
\captionsetup{justification=centering}
\caption{Logistic models}
\label{tab:regr}
\end{subtable}
\caption{Descriptive statistics and estimated logistic models for CUN data. \label{tab:descregr}}
\end{table}

As unmeasured outcome-related confounders (such as genetic factors) may be present, previous results are used to build estimated identification bounds and uncertainty intervals for log odds-ratio NDEs, NIEs and TEs. More precisely, we follow the strategy outlined in Section~\ref{sec:est}, fitting a logistic regression model for the outcome and for the mediator and obtaining - for any desired $(x,x^{\star},c)$ combination - the estimated matrices $\hat{\bm{\Sigma}}_{x,x^{\star}\mid c}$ and $\hat{\bm{D}}_{x,x^{\star}\mid c}$ from these models. The fitted models are reported in Table~\ref{tab:regr}. In line with the above literature, the coefficients of $X$ and $M$ in the outcome model as well as that of $X$ in the mediator model are positive and highly significant. In both models, BMI and Gender are included as covariates as they are relevant predictors of the exposure $X$ as shown by a dedicated linear model (not reported). In fact, since no assumption on outcome-related confounding is made, these variables could in principle be omitted from the outcome model. Nonetheless, adjusting the effect of $X$ and $M$ on $Y$ for BMI and Gender seems in order from a subject-matter standpoint, and therefore we opt to keep them as control variables, although their estimated effect is not highly significant. As for Age, the variable is not included in the models as its effect is aliased with that of $X$, and, consequently, the resulting estimated coefficients are not significant. This is not surprising, in that $X$ is a cumulative measure of lifetime tobacco consumption. In both logistic models, no significant interaction or nonlinear effect was found.

\begin{figure}[p]
\captionsetup[subfigure]{justification=centering}
\centering{
    \begin{subfigure}{0.45\textwidth}
        \includegraphics[width=\hsize]{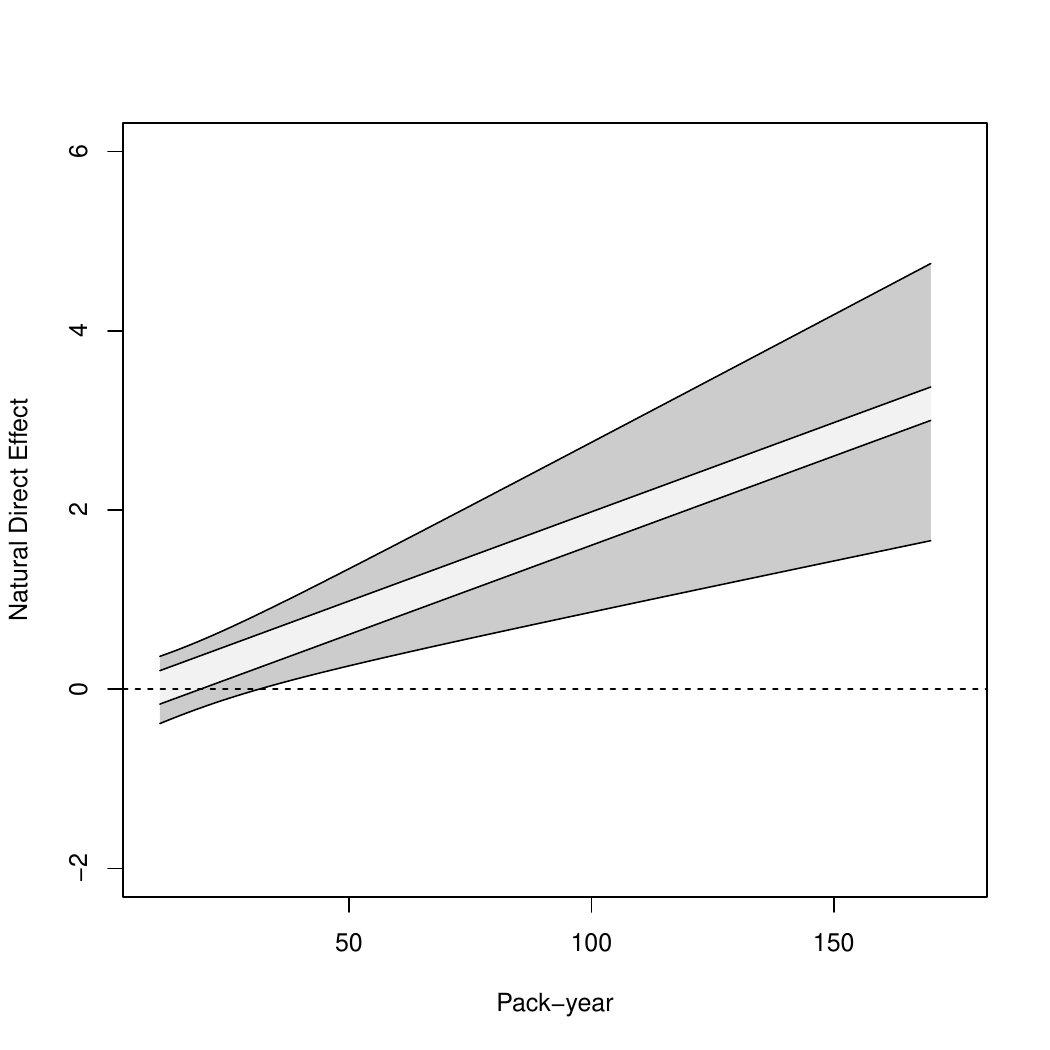}
        \caption{NDE: Female, BMI=25.05}
%        \label{fig.SICAPI}
    \end{subfigure} \quad
     \begin{subfigure}{0.45\textwidth}
        \includegraphics[width=\hsize]{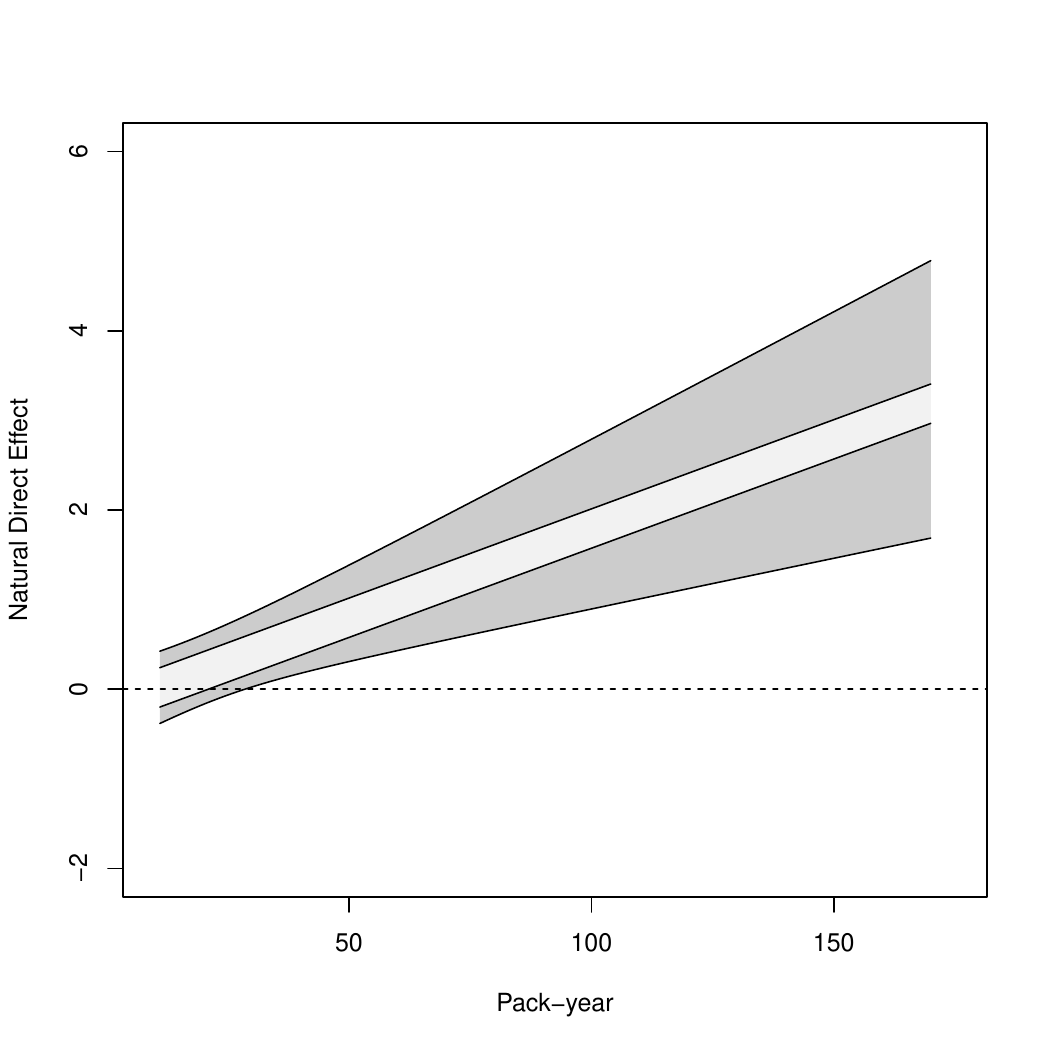}
        \caption{NDE: Male, BMI=28.50}
%        \label{fig.SICAPI}
    \end{subfigure} \\
     \begin{subfigure}{0.45\textwidth}
        \includegraphics[width=\hsize]{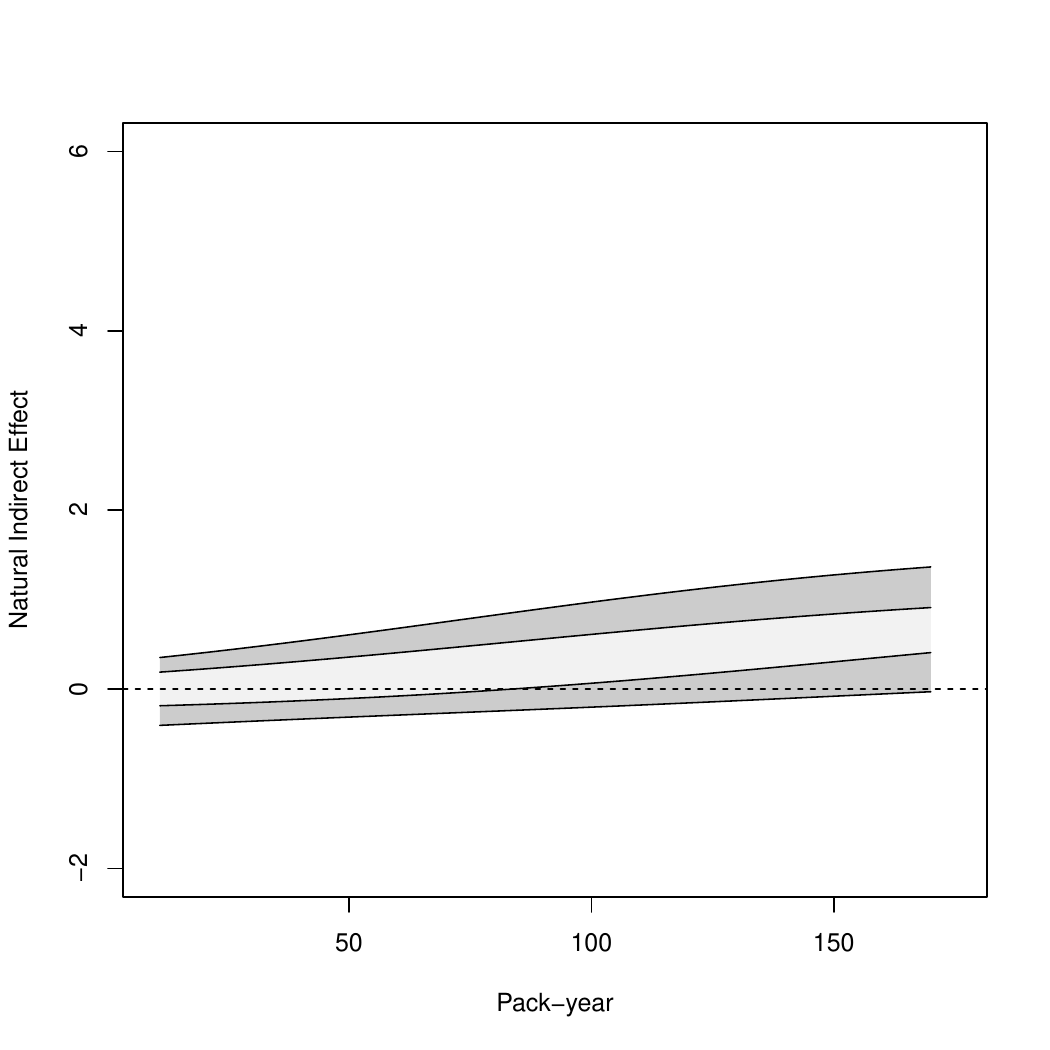}
        \caption{NIE: Female, BMI=25.05}
%        \label{fig.SICAPI}
    \end{subfigure} \quad
     \begin{subfigure}{0.45\textwidth}
        \includegraphics[width=\hsize]{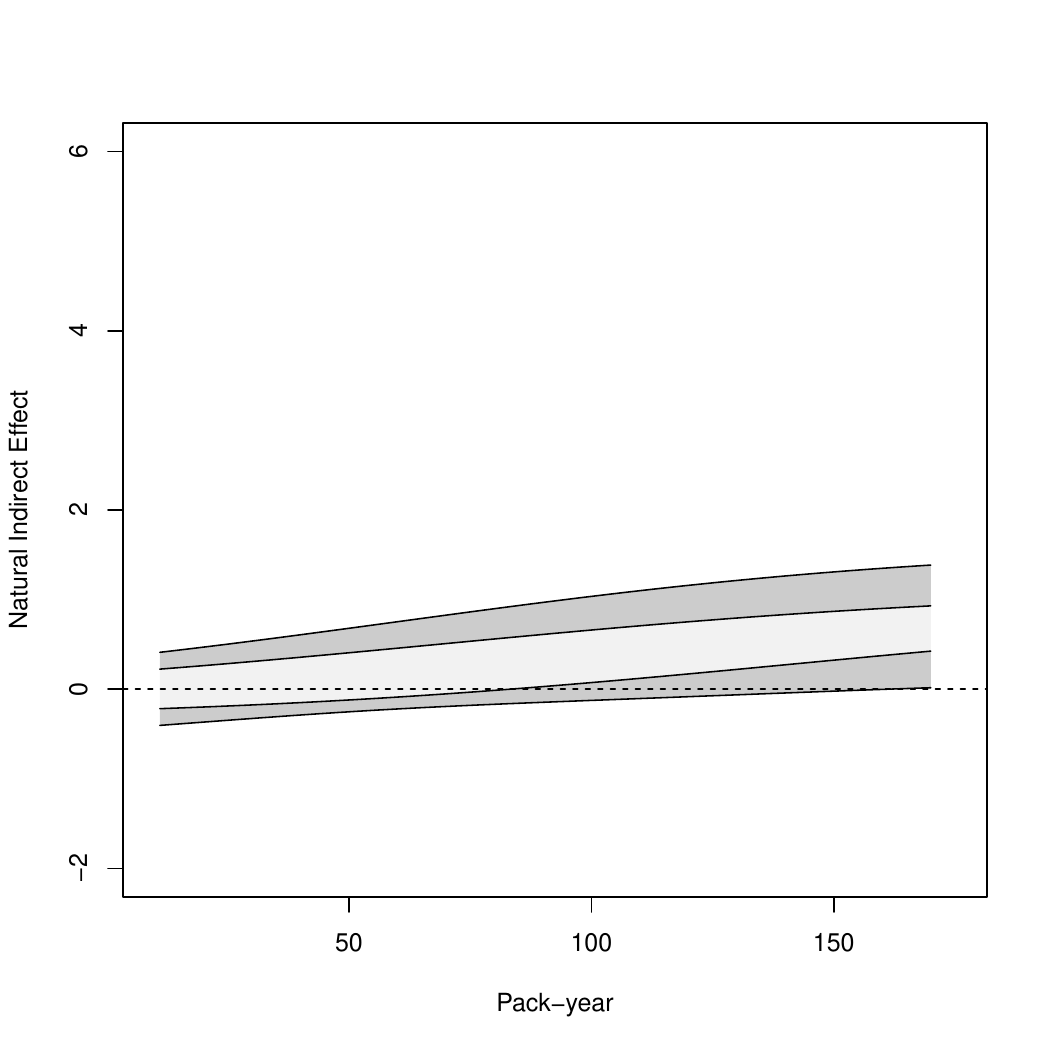}
        \caption{NIE: Male, BMI=28.50}
%        \label{fig.SICAPI}
    \end{subfigure} \\   
}
\caption{Estimated identification bounds (in white) and uncertainty intervals (in grey) for log odds-ratio NDEs and NIEs in the CUN lung cancer study data. The $x$-axis reports the active level, $x$, which is always contrasted to the reference level $x^{\star}=10$ pack-years. Effects are conditional on gender and BMI, which is set to the average of females (25.05 kg/m\textsuperscript{2}) and males (28.50 kg/m\textsuperscript{2}), respectively.}
\label{fig:neffs}
\end{figure}

Figure~\ref{fig:neffs} depicts estimated identification bounds (in white) and uncertainty intervals (in grey) for some log odds-ratio NDEs (top plots) and NIEs (bottom plots) constructed from the above regression models. In particular, bounds/intervals are computed separately for females (left plots) and males (right plots), with the BMI set to the respective average values (25.05 and 28.50 kg/m\textsuperscript{2}). In all plots, the \emph{x}-axis contains the active treatment level, $x$, whereas the reference level is set to $x^{\star}=10$ pack-years, that is, the lowest smoking intensity among enrolled patients. The four plots show positive and quite large NDEs, with uncertainty intervals not including 0 for relatively low values of $x$. In contrast, NIEs are smaller, and their uncertainty intervals always contain 0. As a consequence, plots for the TEs (not shown) are very similar to those of the NDEs.

%%%%%%%%%%%%%%%%%%%%%%%%%%%

\section{Discussion}\label{sec:concl}
In a setting with a binary outcome and a binary mediator, we have proposed identification bounds and uncertainty intervals for log odds-ratio natural mediation effects under a restricted set of assumptions. In particular, no assumption on the absence of unmeasured confounding for the outcome-exposure and outcome-mediator relationships is made. Furthermore, the Cross-World Independence (CWI) assumption usually needed for point identification is replaced by a weaker condition termed Partially Constant Cross-World Dependence (PC-CWD). Our strategy is semi-parametric in that it invokes a logistic regression model for both the mediator and the outcome, but the functional form in the right-hand side of these models governing the dependence on the explanatory variables is arbitrary. In other words, for each model the predictor obtained from the explanatory variables does not have to be linear in the parameters.

The derivations are applied to data from a lung cancer screening program in order to investigate whether the effect of smoking intensity on lung cancer onset is mediated by the presence of some kind of emphysema, a pulmonary disease induced by smoking and associated to lung cancer even adjusting for tobacco exposure. In this context, the presence of unmeasured confounders of the exposure-outcome and mediator-outcome relation (such as genetic factors) cannot be excluded. Results show a quite strong and significant natural direct effect, whereas the natural indirect effect is weaker and not significant. These effects are conditional on gender and Body Mass Index (BMI), which are potential confounders for the exposure-mediator relationship.

Because of the adoption of logistic models, our partial identification strategy is naturally suited to effects on the (log) odds-ratio scale. Moving to other scales (e.g., risk ratio, risk difference) is more problematic. This is due to the fact that effects on these alternative scales cannot be expressed as a function of (log) odds-ratio effects without specifying the underlying counterfactual probabilities, which are not point-identified under the reduced assumption set. With this regard, specific approaches for scales other than (log) odds-ratio would represent an interesting stream of future research. Another possible extension could consist in embedding the alternative identification strategies of~\cite{Pearl2014}, mentioned in Section~\ref{subsec:notass}. While differentiating the covariate sets is relatively straightforward (see for example the discussion in Section~\ref{sec:appl} with respect to the application at hand), challenges remain in relation to the inclusion of identification strategies different from covariate adjustment, such as those based on the front-door criterion.

%%%%%%%%%%%%%%%%%%%%%%%%%%%

\appendix

%%%%%%%%%%%%%%%%%%%%%%%%%%%

\section{Natural effect model for point identification}\label{app:proofnem}
Conditioning on covariates $C$ is omitted for brevity. Under the semi-parametric models~\eqref{py} and~\eqref{pm} and assumptions \emph{i)-iv)}, from first principles in probability we have
\[
\textup{logit}\{P(Y(x,M(x^{\star}))=1\mid M(x^{\star})=m)\} = \log\frac{P(M(x^{\star})=m\mid Y(x,M(x^{\star}))=1)}{P(M(x^{\star})=m\mid Y(x,M(x^{\star}))=0)} + \textup{logit}\{P(Y(x,M(x^{\star}))=1)\},
\]
from which we can write
\begin{equation}\label{eq:proofnemcwi}
\begin{split}
\textup{logit}\{P(Y(x,M(x^{\star}))=1)\} &= \textup{logit}\{P(Y(x,M(x^{\star}))=1\mid M(x^{\star})=m)\}+\log\frac{P(M(x^{\star})=m\mid Y(x,M(x^{\star}))=0)}{P(M(x^{\star})=m\mid Y(x,M(x^{\star}))=1)} \\
&= \textup{logit}\{P(Y(x,m)=1\mid M(x^{\star})=m)\}+\log\frac{P(M(x^{\star})=m\mid Y(x,M(x^{\star}))=0)}{P(M(x^{\star})=m\mid Y(x,M(x^{\star}))=1)} \\
&= \textup{logit}\{P(Y(x,m)=1)\}+\log\frac{P(M(x^{\star})=m\mid Y(x,M(x^{\star}))=0)}{P(M(x^{\star})=m\mid Y(x,M(x^{\star}))=1)} \\
&= r_{\beta}(x,m)+\log\frac{P(M(x^{\star})=m\mid Y(x,M(x^{\star}))=0)}{P(M(x^{\star})=m\mid Y(x,M(x^{\star}))=1)}. \\
\end{split}
\end{equation}
In detail, the second equality follows from the fact that, conditional on the event $M(x^{\star})=m$, the random variable $Y(x,M(x^{\star}))$ can be replaced by $Y(x,m)$, the third is implied by the CWI assumption \emph{(iv)}, while the fourth follows from \emph{(i)} and \emph{(ii)} paired with consistency and model~\eqref{py}. In order to identify the counterfactual term
\[
\log\frac{P(M(x^{\star})=m\mid Y(x,M(x^{\star}))=0)}{P(M(x^{\star})=m\mid Y(x,M(x^{\star}))=1)},
\]
one can make repeated use of the above scheme. Specifically, we have
\begin{equation}\label{eq:proofnem2cwi}
\begin{split}
\textup{logit}\{P(M(x^{\star})=1\mid Y(x,M(x^{\star}))=y)\} &= \log\frac{P(Y(x,M(x^{\star}))=y \mid M(x^{\star})=1)}{P(Y(x,M(x^{\star}))=y \mid M(x^{\star})=0)}+\textup{logit}\{P(M(x^{\star})=1)\}\\
&= \log\frac{P(Y(x,M(x^{\star}))=y \mid M(x^{\star})=1)}{P(Y(x,M(x^{\star}))=y \mid M(x^{\star})=0)}+ r_{\gamma}(x^{\star})\\
&= \log\frac{P(Y(x,1)=y \mid M(x^{\star})=1)}{P(Y(x,0)=y \mid M(x^{\star})=0)}+ r_{\gamma}(x^{\star})\\
&= \log\frac{P(Y(x,1)=y)}{P(Y(x,0)=y)}+ r_{\gamma}(x^{\star})\\
&= \log\Biggl\{\frac{\exp\{y\cdot r_{\beta}(x,1)\}}{\exp\{y\cdot r_{\beta}(x,0)\}}\cdot\frac{1+\exp\{r_{\beta}(x,0)\}}{1+\exp\{r_{\beta}(x,1)\}}\Biggr\}+ r_{\gamma}(x^{\star})\\
&= y\{r_{\beta}(x,1)-r_{\beta}(x,0)\}+\log\frac{1+\exp\{r_{\beta}(x,0)\}}{1+\exp\{r_{\beta}(x,1)\}}+r_{\gamma}(x^{\star}) \\
&\equiv g_y(x,x^{\star}).
\end{split}
\end{equation}
Again, the first equality follows from first principles, while the second is due to \emph{(iii)} paired with consistency and model~\eqref{pm}. The third equality is implied by the same argument implying the second equality of the previous chain, while the fourth equality derives from CWI, and the fifth one from \emph{(i)} and \emph{(ii)} paired with consistency and model~\eqref{py}.

Finally, one can compute the right-hand side of~\eqref{eq:proofnemcwi} in $m=0$ and replace the risk ratio with the expression deriving from~\eqref{eq:proofnem2cwi}. This leads to
\[
\begin{split}
\textup{logit}\{P(Y(x,M(x^{\star}))=1)\} &= r_{\beta}(x,0)+\log\frac{P(M(x^{\star})=0\mid Y(x,M(x^{\star}))=0)}{P(M(x^{\star})=0\mid Y(x,M(x^{\star}))=1)} \\
&= r_{\beta}(x,0)+\log\frac{1+\exp\{g_1(x,x^{\star})\}}{1+\exp\{g_0(x,x^{\star})\}},
\end{split}
\]
thereby proving, up to the inclusion of covariates, Equation~\eqref{eq:nem}.

\section{Semi-parametric bounds for $p^{(\psi)}_{x,x^{\star}\mid c}$}\label{app:boundsp}
To derive the bounds in~\eqref{eq:boundsp1} and~\eqref{eq:boundsp2}, it is convenient to examine the dependence on the shift parameter $\psi$ of
\[
\tilde{g}_0(\psi,x,x^{\star};c) = \log\frac{1+\exp\{\psi+r_{\beta}(x,0;c)\}}{1+\exp\{\psi+r_{\beta}(x,1;c)\}} + r_{\gamma}(x^{\star};c),
\]
taking the other arguments as fixed. From first mathematical principles, it holds that
\[
\lim_{\psi\to-\infty} \tilde{g}_0(\psi,x,x^{\star};c)  = r_{\gamma}(x^{\star};c) \qquad\qquad\qquad\qquad \lim_{\psi\to+\infty} \tilde{g}_0(\psi,x,x^{\star};c)  = r_{\gamma}(x^{\star};c) - \delta_{\beta}(x;c).
\]
Also, $\tilde{g}_0(\psi,x,x^{\star};c)$ is a monotonic function of $\psi$. This fact can be proven by noticing that the derivative
\[
\frac{\partial}{\partial \psi} \tilde{g}_0(\psi,x,x^{\star};c) = \frac{\exp\{\psi+r_{\beta}(x,0;c)\}-\exp\{\psi+r_{\beta}(x,1;c)\}}{[1+\exp\{\psi+r_{\beta}(x,0;c)\}][1+\exp\{\psi+r_{\beta}(x,1;c)\}]}
\]
is always non-null if one rules out extreme settings where $\delta(x;c)=0$. In these settings, the indirect effect is null, and the whole decomposition is not meaningful anymore. Thus, it holds that
\[
r_{\gamma}(x^{\star};c) - \delta_{\beta}(x;c) \leq \tilde{g}_0(\psi,x,x^{\star};c) \leq r_{\gamma}(x^{\star};c)
\]
if $\delta_{\beta}(x;c) > 0$ and
\[
r_{\gamma}(x^{\star};c) \leq \tilde{g}_0(\psi,x,x^{\star};c) \leq r_{\gamma}(x^{\star};c) - \delta_{\beta}(x;c)
\]
if $\delta_{\beta}(x;c) < 0$, from which~\eqref{eq:boundsp1} and~\eqref{eq:boundsp2} immediately descend.

%%%%%%%%%%%%%

\section{Derivative matrix for Delta method}\label{app:delta}
The $\bm{D}_{x,x^{\star}\mid c}$ matrix in~\eqref{eq:v0} can be obtained by column-binding the four derivative vectors associated to the elements of $\bm{\tau}_{x,x^{\star}\mid c}$. In practice, we have
\[
\bm{D}_{x,x^{\star}\mid c} = \frac{\partial\bm{\tau}^\top_{x,x^{\star}\mid c}}{\partial\bm{\theta}_{x,x^{\star}\mid c}} = \begin{pmatrix} \bm{d}(\ell)_{x,x^{\star}\mid c}^{\textup{NDE}},\bm{d}(u)_{x,x^{\star}\mid c}^{\textup{NDE}},\bm{d}(\ell)_{x,x^{\star}\mid c}^{\textup{NIE}},\bm{d}(u)_{x,x^{\star}\mid c}^{\textup{NIE}}\end{pmatrix},
\] 
where, letting $\textup{expit}(\cdot)=\exp(\cdot)/\{1+\exp(\cdot)\}$, 
\[
\bm{d}(\ell)_{x,x^{\star}\mid c}^{\textup{NDE}} = \begin{pmatrix}[1.5] 1-\textup{expit}\{r_{\gamma}(x^{\star};c)-\delta_{\beta}(x;c)\} \\
\textup{expit}\{r_{\gamma}(x^{\star};c)+\delta_{\beta}(x^{\star};c)\}-1 \\
\textup{expit}\{r_{\gamma}(x^{\star};c)-\delta_{\beta}(x;c)\} \\
-\textup{expit}\{r_{\gamma}(x^{\star};c)-\delta_{\beta}(x^{\star};c)\} \\
0 \\
2\cdot\textup{expit}\{r_{\gamma}(x^{\star};c)\} - \textup{expit}\{r_{\gamma}(x^{\star};c)-\delta_{\beta}(x;c)\} - \textup{expit}\{r_{\gamma}(x^{\star};c)+\delta_{\beta}(x^{\star};c)\} 
\end{pmatrix},
\]
\[
\bm{d}(u)_{x,x^{\star}\mid c}^{\textup{NDE}} = \begin{pmatrix}[1.5] 1-\textup{expit}\{r_{\gamma}(x^{\star};c)+\delta_{\beta}(x;c)\} \\
\textup{expit}\{r_{\gamma}(x^{\star};c)-\delta_{\beta}(x^{\star};c)\}-1 \\
\textup{expit}\{r_{\gamma}(x^{\star};c)+\delta_{\beta}(x;c)\} \\
-\textup{expit}\{r_{\gamma}(x^{\star};c)-\delta_{\beta}(x^{\star};c)\} \\
0 \\
-2\cdot\textup{expit}\{r_{\gamma}(x^{\star};c)\} + \textup{expit}\{r_{\gamma}(x^{\star};c)+\delta_{\beta}(x;c)\} + \textup{expit}\{r_{\gamma}(x^{\star};c)-\delta_{\beta}(x^{\star};c)\} 
\end{pmatrix}
\]
and
\[
\bm{d}(\ell)_{x,x^{\star}\mid c}^{\textup{NIE}} = \begin{pmatrix}[1.5] \textup{expit}\{r_{\gamma}(x^{\star};c)+\delta_{\beta}(x;c)\}-\textup{expit}\{r_{\gamma}(x;c)-\delta_{\beta}(x;c)\} \\
0 \\
\textup{expit}\{r_{\gamma}(x;c)-\delta_{\beta}(x;c)\} - \textup{expit}\{r_{\gamma}(x^{\star};c)+\delta_{\beta}(x;c)\} \\
0 \\
\textup{expit}\{r_{\gamma}(x;c)\} - \textup{expit}\{r_{\gamma}(x;c)-\delta_{\beta}(x;c)\} \\
\textup{expit}\{r_{\gamma}(x^{\star};c)\} - \textup{expit}\{r_{\gamma}(x^{\star};c)+\delta_{\beta}(x;c)\} \\
\end{pmatrix},
\]
whereas
\[
\bm{d}(u)_{x,x^{\star}\mid c}^{\textup{NIE}} = \begin{pmatrix}[1.5] \textup{expit}\{r_{\gamma}(x^{\star};c)-\delta_{\beta}(x;c)\}-\textup{expit}\{r_{\gamma}(x;c)+\delta_{\beta}(x;c)\} \\
0 \\
\textup{expit}\{r_{\gamma}(x;c)+\delta_{\beta}(x;c)\} - \textup{expit}\{r_{\gamma}(x^{\star};c)-\delta_{\beta}(x;c)\} \\
0 \\
\textup{expit}\{r_{\gamma}(x;c)+\delta_{\beta}(x;c)\} - \textup{expit}\{r_{\gamma}(x;c)\} \\
\textup{expit}\{r_{\gamma}(x^{\star};c)-\delta_{\beta}(x;c)\} - \textup{expit}\{r_{\gamma}(x^{\star};c)\} \\
\end{pmatrix}.
\]

%%%%%%%%%%%%%%%%%%%%%%%%%%%

\bibliographystyle{abbrvnat}      % basic style, author-year citations
\setcitestyle{abbrvnat,open={(},close={)}} %Citation-related commands
% \bibliography{BiblioSens}

\end{document}